\documentclass[10pt]{article}
\usepackage{amssymb}
\usepackage{graphics}

\newtheorem{theorem}{Theorem}[section]

\newtheorem{definition}[theorem]{Definition}
\newtheorem{deff}[theorem]{Definition}

\newtheorem{claim}[theorem]{Claim}
\newtheorem{lemma}[theorem]{Lemma}

\newtheorem{corollary}[theorem]{Corollary}

\newcommand{\qedsymb}{\hfill{\rule{2mm}{2mm}}}
\newenvironment{proof}[1][]{\begin{trivlist}
\item[\hspace{\labelsep}{\bf\noindent Proof#1:\/}] }{\qedsymb\end{trivlist}}

\def\calP{{\cal P}}

\def\R{\mathbb{R}}
\def\C{\mathbb{C}}

\def\mod{{\rm{mod}}}
\def\vol{{\rm{vol}}}
\def\diam{{\rm{diam}}}
\def\tr{{\rm{tr}}}

\def\ra{\rangle}
\def\la{\langle}

\newcommand\ket[1]{{ |{#1} \rangle }}
\newcommand\bra[1]{{ \langle {#1} | }}

\newcommand{\onote}[1]{}
\newcommand{\dnote}[1]{}

\renewcommand{\epsilon}{\varepsilon}

\begin{document}

\title{\bf A Lattice Problem in Quantum NP}

\author{
 Dorit Aharonov\footnote{School of Computer Science and Engineering, The Hebrew University, Jerusalem, Israel. doria@cs.huji.ac.il. Research
supported by ISF grant 032-9738.}
 \and
 Oded Regev \footnote{Institute for Advanced Study, Princeton, NJ. odedr@ias.edu. Research
supported by NSF grant CCR-9987845.} }


\maketitle

\begin{abstract}
We consider $coGapSVP_{\sqrt{n}}$, a gap version of the shortest vector in a lattice problem. This problem is known to
be in $AM\cap coNP$ but is not known to be in $NP$ or in $MA$. We prove that it lies inside $QMA$, the quantum analogue of
$NP$. This is the first non-trivial upper bound on the quantum complexity of a lattice problem.

 The proof relies on two novel
ideas. First, we give a new characterization of $QMA$, called $QMA+$.
Working with the $QMA+$ formulation allows us
to circumvent a problem which arises commonly in the context of $QMA$:
 the prover might use entanglement between different
copies of the same state in order to cheat.
The second idea involves using estimations of
autocorrelation functions for verification. We make the important
observation that autocorrelation functions are positive definite
functions and using
properties of such functions we severely
restrict the prover's possibility to cheat.
We hope that these ideas will lead to
further developments in the field.
\end{abstract}

\section{Introduction}

The field of quantum algorithms has witnessed several important
 results (e.g., \cite{hallgren, watrous, legendre, santha, ChildsSpeedupWalk}) in
the last decade, since the breakthrough discovery of Shor's quantum algorithm for factoring and discrete logarithm in
1994 \cite{ShorFactor}. Despite these important developments,
two problems in particular had little progress in terms of quantum algorithms: graph isomorphism (GI), and gap
versions of lattice problems such as the shortest vector in the lattice problem (GapSVP) and the closest vector in the
lattice problem (GapCVP).

To understand why these problems are interesting in the context of quantum computation, let us first recall their
definitions and what is known about them classically. Graph isomorphism is the problem of deciding whether two given
graphs can be permuted one to the other. It is known to be in $NP \cap coAM$ \cite{GMWGraphIsomorphism} and therefore,
it is not NP complete unless the polynomial hierarchy collapses. $GapSVP_{\beta(n)}$ is the problem of deciding whether
the shortest vector in a given $n$-dimensional lattice $L$ is shorter than $1$ or longer than $\beta(n)$.
$GapCVP_{\beta(n)}$ is the following problem: Given a lattice and a vector $v$, decide whether $d(v,L)\le 1$ or
$d(v,L)> \beta(n)$ where $d(v,L)$ is the minimal distance between $v$ and any point in $L$. Both problems have
important cryptographic applications \cite{MicciancioBook}. Regarding their complexity, it is easy to see that they
both lie in $NP$ for any $\beta(n)\ge 1$. The results of Lagarias et al. \cite{LagariasKZ} imply that when
$\beta(n)=\Omega(n)$, both problems are in $coNP$. For $\beta(n)=\Omega(\sqrt{n/\log(n)})$ these lattice problems are
not known to be in $coNP$ but as shown in \cite{GG}, they are in $coAM$ (and in fact in the class Statistical
Zero Knowledge). This implies that for $\beta(n)=\Omega(\sqrt{n/\log(n)})$ the problems are not NP complete unless the
polynomial hierarchy collapses.

The fact that the graph isomorphism problem and the two lattice problem with the above parameters
are very unlikely to be NP complete, and that they possess a lot of structure, raised
the hope that quantum computers might be able to solve them more efficiently than classical computers. Despite many
attempts,  so far all that is known in terms of the quantum complexity of these problems are reductions to problems for
which quantum algorithms are also not known \cite{AharonovTaShma, RegevQuantumLattices, EttingerHoyer}, and negative
results regarding possible approaches \cite{vazirani, RegevPKE}. Progress in designing an algorithm for one of these
problems is the holy grail of quantum algorithmic theory.

In light of the difficulty of finding efficient algorithms for these
 problems, a weaker question attracted attention: can any quantum upper bound be given
on these problems, which does not follow trivially from the classical upper bounds? Regarding graph isomorphism, which
is known to be in coAM, the natural question to ask is whether it is in coQMA, the quantum analog of coNP. It is more
natural to speak in this context, and in the rest of the paper, about the {\it complements} of the problems we
described, and so the question is whether the graph {\it non}-isomorphism (GNI) problem lies inside QMA. QMA  can be
viewed as the quantum analog of NP, and was recently studied in various papers
\cite{KitaevBook,watrous,DoritQMASurvey,KempeRegev,ShpilkaRaz}. Strictly speaking, QMA is actually the analog of Merlin
Arthur, the probabilistic version of NP, since in the quantum world it is more natural to consider probabilistic
classes. Attempts to prove that GNI is in QMA  have so far failed. As for lattice problems, since $NP \subseteq QMA$,
it follows from the classical result \cite{LagariasKZ} that if $\beta(n)=\Omega(n)$ the complements of the problems we
described, namely $coGapCVP$ and $coGapSVP$, lie in $QMA$. The interesting question, however, is whether these problems are
still in $QMA$ for lower gaps, such as $\beta(n)=\Omega(\sqrt{n})$. Notice that this does not follow from the classical
results.

\subsection{Results}

In this paper we solve the question of containment in QMA for one of the aforementioned problems. This is the first non
trivial quantum upper bound for a lattice problem.

\begin{theorem}\label{thm:main}
The problem $coGapSVP_{c \sqrt{n}}$ is in $QMA$ for some constant $c>0$.
\end{theorem}

One of the new ideas in the proof of Theorem \ref{thm:main} is the important connection between quantum estimations of
inner products, or autocorrelation estimates, and properties of positive definite functions. The technique of using
positive definite functions to analyze quantum protocols is likely to prove useful in other contexts, due to its
generality: the property of positive definiteness applies to autocorrelation functions over any group, and not only
over $\R^n$ as in our case.

Another important issue in the proof Theorem \ref{thm:main} is a problem that arises commonly in the analysis of QMA
protocols. Namely, in certain situations, we would like to repeat
a test on several copies of the witness but the
prover might use entanglement between the copies in order to cheat.
We circumvent this problem
 by giving a new characterization of $QMA$, named $QMA+$. We start by proving that
indeed $QMA=QMA+$ and then, using this new characterization,
we prove the soundness of our protocol.

\subsection{Open Questions}

Hopefully, both the new characterization of QMA and the new technique of verification using positive definite functions
will help in proving that other important problems such as $GNI$ and $coGapCVP_{\sqrt{n}}$ lie in $QMA$.

In more generality, in this work we gain a better understanding of the class QMA and the techniques used to analyze it.
We hope that this work will lead to an even better understanding of this important class.  Understanding classical NP
led to a few of the most important results in theoretical computer science, including PCP and hardness of
approximation. A few indications that QMA is fundamental for quantum computation have already been given in
\cite{DoritQMASurvey, adiabatic}.

Our results might also lead to progress in terms of quantum algorithms for lattice problems. In this context, it is
interesting to consider Theorem \ref{thm:main} in light of a recent paper by Aharonov and
 Ta-Shma \cite{AharonovTaShma}. \cite{AharonovTaShma} showed that if the state
we use as the quantum witness in the QMA protocol
can be {\it generated} efficiently, it can be used to provide a $BQP$
algorithm for the lattice problem.
The result we present here shows that certain properties of the state of \cite{AharonovTaShma}
can be {\it verified} efficiently, which might
be a stepping stone towards
understanding how to generate the state efficiently, thus providing an
efficient algorithm for the lattice problem.

Finally, we mention that
similar techniques to the one used in the proof of $QMA=QMA+$,
 might also prove useful in other contexts, for example
for proving security
of quantum cryptographic protocols.


\subsection{Outline of the Paper} The paper starts with an overview of the proof. We continue with preliminaries
 in Section \ref{sec:prelim}.
The proof of Theorem \ref{thm:main} is obtained by
combining three theorems. The proof of each of the theorems
 is independent and is presented in a
separate section. First, in Section \ref{sec:qma+} we define the class QMA+ and show that it is equal to QMA. Then,
in Section \ref{sec:cvp}, we show that $coGapCVP'$, a version of $coGapCVP$,
is in QMA+. Finally, in Section \ref{sec:svptocvp} we show that if
$coGapCVP'$ is in QMA then so is $coGapSVP$.

\section{Overview of the Proof}

Assume we are given a witness which we would like to verify. Usually, we apply a certain unitary transformation and
measure the output qubit. If the witness is correct, the outcome should be $1$. Hence, we reject if the outcome is $0$.
Consider, however, a situation where our unitary transformation is such that for the correct witness the outcome is $1$
with probability $p$, for some $p>0$. Thus, it is natural to consider the following stronger test: we apply a unitary
transformation and accept if the {\em probability} of measuring $1$ is close to some number $p$. We call a verifier
that performs such tests a super-verifier and denote the corresponding class by $QMA+$. Our first theorem is
\begin{theorem}\label{thm:qma+}
$QMA=QMA+$
\end{theorem}

Showing that $QMA$ is contained in $QMA+$ is easy; essentially, the super-verifier can say that the probability of
measuring $1$ should be close to $p=1$. The other direction is more interesting. Given a super-verifier we can
construct a verifier that accepts a witness which is composed of many copies of the original witness. The verifier can
then apply the unitary transformation to each one of the copies and measure the results. Finally, it can compute the
fraction of times $1$ was measured and check if it is close to $p$. Indeed, if the prover does not cheat and sends many
copies of the original witness we should measure $1$ in around a $p$ fraction of the measurements. However, it seems
that the prover might be able to cheat by using entanglement between the
different copies. Using the Markov inequality, we show that this is impossible.

Next, we show
\begin{theorem}\label{thm:gapcvp'}
The problem $coGapCVP'_{c \sqrt{n}}$ is in $QMA+$ for some constant $c>0$.
\end{theorem}
 $coGapCVP'_{\beta(n)}$ is a variant of $coGapCVP_{\beta(n)}$ where we are given the additional promise that the
shortest vector in $L$ is longer than $\beta(n)$. The proof of this theorem is very involved, but the idea is as
follows.

\begin{figure}
 \centering \resizebox{3in}{2in}{\includegraphics[40,246][747,736]{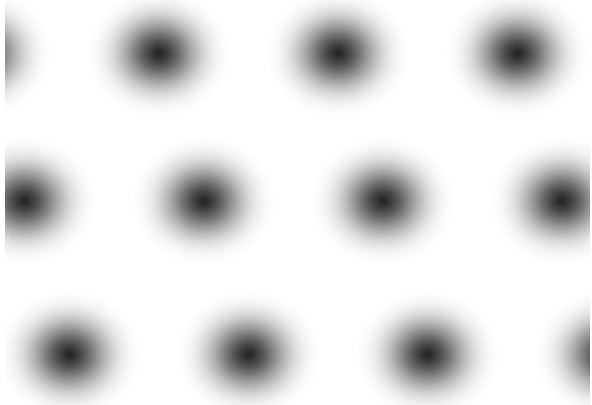}}
 \caption{The quantum witness} \label{fig:witness}
\end{figure}

The correct quantum witness $|\xi\ra$ for $coGapCVP'$, i.e., the witness in case $v$ is far from the lattice, is
defined as follows (a similar state appears in \cite{AharonovTaShma} which can be seen as the quantum analogue of the
probability distribution of \cite{GG}). Consider the `probability distribution' obtained by choosing a random lattice
point and adding to it a Gaussian of radius $\sqrt{n}$. We define $|\xi\ra$ as the superposition corresponding to this
probability distribution. See Figure \ref{fig:witness}. Actually, the state $|\xi\ra$ cannot be defined as above, since
we cannot represent a point in $\R^n$ with infinite precision, so we need to work over a very fine grid. Moreover, the
number of grid points in $\R^n$ is infinite. Hence, we restrict the state to grid points inside the basic
parallelepiped of the lattice. We will define this formally later; it is best to keep in mind the continuous picture.

Given this superposition, for some constant $c$, solving $coGapCVP'_{c \sqrt{n}}$ (and in fact also
$coGapCVP_{c \sqrt{n}}$) is easy: it is done by estimating the inner product of the above state with the same state
shifted by $v$. If $d(v,L)\ge c\sqrt{n}$ then the inner product is almost zero since the Gaussians and their shifted
version do not intersect. If $d(v,L) \le 1$, the inner product is large since the two states are almost the same.
 To show containment in $QMA+$, we will use
this state as the correct witness. Hence, it remains to show how a super-verifier can verify that the prover is not cheating.
Cheating in this context means that $d(v,L) \le 1$ but the prover claims that $d(v,L) \ge c\sqrt{n}$ and sends some
witness which is not necessarily the correct witness.

We now define the verification process. Define $h(x)$ to be the real part of the
inner product of the given witness state with itself shifted by $x$. We call $h$ the autocorrelation function of the
witness. It is a function from $\R^n$ to $\R$ such that $h(0)=1$.
We define $g$ to be the same, for the correct witness $|\xi\ra$.
An important property of $h$ is that for any $x$, there exists a quantum circuit
whose probability of outputting $1$ is directly related to $h(x)$.
Hence, since a super-verifier can check the probability of outputting $1$, it can
effectively check that $h(x)$ is close to some value.
Since we expect to see the correct witness, we construct a super-verifier
that checks that $h(x)$ is close to $g(x)$ for some vectors $x$. More precisely,
with probability half the super-verifier chooses the vector $v$ and otherwise
it randomly chooses a polynomially short vector.

In order to complete the description of the super-verifier, we have to show
that it can compute $g(x)$ for the points chosen above.
Later in the paper we analyze the function $g$ and it turns out to have a familiar form: it is very
close to a periodic Gaussian, like the one shown in Figure \ref{fig:witness}.
Therefore, $g(v)$ is approximately zero since $v$ is far from the lattice and
$g(x)$ for short vectors $x$ has the form $e^{-\|x\|^2}$.
In both cases, the super-verifier knows the value of $g$ and can
therefore perform the verification procedure described above.
We remark that analyzing $g$ involves some technical calculations;
It is here that we need the assumption that the shortest vector in the
lattice is large, so that the Gaussians are well separated and do not
interfere with each other.

The proof of soundness of this test uses the observation that
autocorrelation functions are necessarily positive definite.
 A function $f$ is positive definite (PD) if for any $k\ge 1$ and any $k$ points $x_1,\ldots,x_k\in \R^n$,
the $k\times k$ matrix $M$ defined by $M_{i,j}=f(x_i-x_j)$ is positive semidefinite. Notice that no matter what witness
the prover gives, the function $h$ must be PD since it is an autocorrelation function. We will complete the proof by
showing that no PD $h$ exists which passes the above test if $d(v,L)\le 1/3$, i.e., no PD function exists which is both
close to $0$ at a vector $v$ whose distance to $L$ is at most $1/3$, and also close to a Gaussian at many randomly
chosen points polynomially close to the origin.

Why doesn't such a PD function exist? Intuitively, our proof relies on certain non-local behaviors of positive definite
functions. Namely, we will show that changing the value of a PD function at even one point affects the function at many
other points. We assume that $h(v)$ is close to $0$ and $d(v,L) \le 1/3$. Let $w$ be a point which is equal to $v$
modulo the lattice (i.e., $w-v\in L$) such that $\|w\| \le 1/3$. Such a point exists since $d(v,L) \le 1/3$. As we will
see later, we can guarantee that $h$ is periodic on the lattice and hence $h(w)=h(v)$ is close to $0$. We start with a
simple property of positive definite functions which can be obtained from using $3\times 3$ matrices in the definition:
if $h(w)$ is close to $0$ then $h(w/2)$ is at most $3/4$ and similarly, $h(w/4)$ is at most $15/16$. By repeating the
argument we derive an upper bound on $h(y)$ where $y=w/2^k$
for some $k>0$. The point $y$ is polynomially close to the
origin and the upper bound is much smaller than the correct
 Gaussian value, $g(y)$. This shows that the super-verifier
can detect a cheating prover by choosing the point $y$. However, the super-verifier does not know where $v$ is relative to the lattice
and therefore he cannot compute $w$ or $y$.
The probability that our randomly chosen point happens to be $y$ is negligible.

Thus, we will have to derive stronger properties of the function $h$. These will be obtained by considering the
positive definite condition with $4\times 4$ matrices. Essentially, we will show that for any point $x$ which
is almost orthogonal to $y$, it cannot be
that $h(x)$, $h(x+y)$ and $h(x-y)$
are all close to their correct
 values $g(x),g(x+y),g(x-y)$. This means that one of the
points in the triple $x,x+y,x-y$
is such that the verifier detects a cheating prover by choosing it.
 Using the fact
that $y$ was chosen to be polynomially short, we will argue that all
 three points in a triple have roughly the same
probability to be chosen by the verifier. Hence, a cheating prover
is caught with non-negligible probability, and the
soundness of the protocol follows.

Curiously, it seems essential in our proof to use Gaussians and not
spheres. This is unlike the classical proof of
\cite{GG} that seems to work both with spheres and with Gaussians.
 Essentially, the difference between the two
distributions is in the behavior of their autocorrelation functions.
For Gaussians, the autocorrelation with a short
vector $x$ behaves like $h(x)\approx 1-c_1\|x\|^2$
while for spheres it behaves like
 $h(x)\approx 1-c_2\|x\|$ where $c_1,c_2$ are some constants.
In the proof, using
properties of positive definite functions obtained from $4\times 4$ matrices, we obtain an upper bound
of the form $h(x)\le 1-c'\|x\|^2$ for some constant $c'>c_1$.
 This yields a contradiction
since $1-c'\|x\|^2 < 1-c_1\|x\|^2$.
 However, if we used spheres, we would not
obtain any contradiction since  $1 - c'\|x\|^2 > 1-c_2\|x\| $
for short vectors $x$.

To complete the proof of Theorem \ref{thm:main}, we need the final theorem:
\begin{theorem}\label{thm:cvptosvp}
For any $\beta=\beta(n)>1$, if $coGapCVP'_{\beta}$ is in $QMA$ then so is
 $coGapSVP_{\beta}$.
\end{theorem}

The proof of this theorem uses an idea similar to \cite{GMSS}. Essentially, an instance of $coGapSVP_{\beta}$ can be
translated into $n$ instances of $coGapCVP'_{\beta}$. If there is no short vector in the original lattice then in all
the $CVP$ instances the target vector is far from the lattice. Otherwise, if there exists a short vector then in at
least one of the $CVP$ instances, the target vector is close to the lattice. Based on this idea, we construct a quantum
verifier for $coGapSVP_{\beta}$. The witness it expects to see is a concatenation of the $n$ witnesses of the
corresponding $coGapCVP'_{\beta}$ problems. It applies a $coGapCVP'_{\beta}$ verifier to each one of the copies and
accepts if and only if they all accept.

\section{Preliminaries}\label{sec:prelim}

\subsection{Definitions}

For $\alpha \in \R$, define $\mu(\alpha)$ as $e^{-\pi \alpha^2}$. For any $x\in \R^n$, we will often denote
$\mu(\|x\|)$ by $\mu(x)$. Let $B_n$ denote the $n$-dimensional unit ball and let $\omega_n$ denote its volume. For a
vector $x\in \R^n$ let $x^\bot$ denote the $n-1$ dimensional subspace orthogonal to $x$. For a vector $x\in \R^n$ and a
subspace $S$ let $P_S(x)$ denote the projection of $x$ on the subspace $S$. We will slightly abuse notation by denoting
the projection of $x$ on the subspace spanned by a vector $v$ as $P_v(x)$.

\subsection{Lattices}
For an introduction to lattices, see \cite{MicciancioBook}. A lattice in $\R^n$ is defined as the set of all integer
combinations of $n$ linearly independent vectors. This set of vectors is known as a basis of the lattice and is not
unique. Given a basis $(v_1,\ldots,v_n)$ of a lattice $L$, the fundamental parallelepiped is defined as
$$\calP(v_1,\ldots,v_n) = \left\{ \sum_{i=1}^n x_i v_i  ~|~  x_i \in [0,1) \right\}.$$
When the basis is clear from the context we will use the notation $\calP(L)$ instead of $\calP(v_1,\ldots,v_n)$. Note
that a lattice has a different fundamental parallelepiped for each possible basis.
For a point $x\in \R^n$ we define $d(x,L)$ as the minimum of $\|x-y\|$ over all $y\in L$.

For a lattice $L=(v_1,\ldots,v_n)$ and a point $x\in \R^n$ we define $x~\mod~L$ as the unique point $y\in
\calP(v_1,\ldots,v_n)$ such that $y-x$ is an integer combination of $v_1,\ldots,v_n$ (see, e.g.,
\cite{Micciancio01hnf}). Notice that a function $f:\calP(L) \rightarrow \C$ can be naturally extended to a function
$f':\R^n \rightarrow \C$ by defining $f'(x):=f(x~\mod~L)$. We will often refer to values of functions outside of
$\calP(L)$, in which case we mean the periodicity above. We will also use, for technical proofs, the notion of a
Voronoi cell of L, denoted $Vor(L)$, which is the set of all points
 in $\R^n$ which are closer to the origin than to
any other lattice point. In addition, $\tau_L(x)$ denotes the unique point $y\in Vor(L)$ such that $y-x\in L$. Notice
that $\|\tau_L(x)\|=d(x,L)$.

\subsection{Shortest and Closest Vector in a lattice}
The shortest (non-zero) vector of $L$ is the vector $x\in L$, such that $\|x\|\neq 0$ and is minimal. The following is
the gap version of the shortest vector problem:

\begin{definition}[coGapSVP]
For any gap parameter $\beta=\beta(n)$ the promise problem $coGapSVP_{\beta}$ is defined as follows. The input is a
basis for a lattice $L$. It is a $YES$ instance if the length of the shortest vector is more than $\beta$. It is a $NO$
instance if the length of the shortest vector is at most $1$.
\end{definition}

We also define the gap version of the closest vector problem and a non-standard variant of it which will be used in
this paper:

\begin{definition}[coGapCVP]
For any gap parameter $\beta=\beta(n)$ the promise problem $coGapCVP_{\beta}$ is defined as follows. The input is a
basis for a lattice $L$ and a vector $v$. It is a $YES$ instance if $d(v,L)>\beta$. It is a $NO$ instance if $d(v,L)\le
1$.
\end{definition}

\begin{definition}[coGapCVP']
For any gap parameter $\beta=\beta(n)$ the promise problem $coGapCVP'_{\beta}$ is defined as follows. The input is a
basis for a lattice $L$ and a vector $v$. It is a $YES$ instance if $d(v,L)>\beta$ and the shortest vector in $L$ is of
length at least $\beta$. It is a $NO$ instance if $d(v,L)\le 1$.
\end{definition}

Each vector in the
input basis $v_1,\ldots,v_n$ is given with polynomially many bits. Without
loss of generality, we assume that the
target vector $v$ is given to us in the form $\sum a_i v_i$ where each $0\le a_i < 1$ is represented by at most
$\ell=poly(n)$ bits.

\subsection{Quantum NP}
We are interested in the quantum analog of the class NP. For an introduction to this class,  the reader is referred to
a recent survey by Aharonov and Naveh \cite{DoritQMASurvey} and to a book by Kitaev,
 Shen and Vyalyi
\cite{KitaevBook}. Strictly speaking, this class is the
quantum analogue of MA, the probabilistic version of NP,
and so it is denoted QMA. It is also sometimes
 denoted BQNP \cite{KitaevBook}.
\begin{definition}[QMA] \label{Def:QMA}
A language $L \in QMA$ if there exists a quantum polynomial
time verifier $V$, polynomials $p,q$, and efficiently computable functions $c,s$,
such that:
\begin{itemize}
 \item $\forall x \in L \quad \exists \rho
\quad \tr(\Pi^{\ket{1}}V \rho V^\dag) \geq c(\frac{1}{|x|})$
 \item $\forall x \notin L \quad \forall \rho
\quad \tr(\Pi^{\ket{1}}V \rho V^\dag) \leq s(\frac{1}{|x|})$, \item $c(\frac{1}{|x|})-s(\frac{1}{|x|})\ge
q(\frac{1}{|x|})$,
\end{itemize}
and the $\rho$'s are density matrices of $p(|x|)$ qubits.
\end{definition}

\subsection{Positive Definite Functions}

\begin{deff}
A $k\times k$ matrix $M$ is positive semidefinite (PSD) if it is Hermitian and for any vector $w\in \C^k$, $w^\dagger M
w$ is real and non-negative.
\end{deff}

The requirement that $M$ is Hermitian is redundant since this is already implied by the requirement that $w^\dagger M
w$ is real for all $w\in \C^k$. The next two claims list some simple properties of positive semidefinite matrices.

\begin{claim}\label{claim:psd_simple}
Let $M,M'$ denote two positive semidefinite matrices. Then the following matrices are also positive semidefinite: $cM$,
$M+M'$, $M^*$ and $Re(M)$ where $c>0$ is real and $Re(M)$
is the matrix obtained by taking the real part of every
entry of $M$.
\end{claim}
\begin{proof}
Clearly, all four matrices are Hermitian. Let $w$ be any vector in $\C^k$. Then, $w^\dagger cM w = c w^\dagger M w \ge
0$ and $w^\dagger (M+M') w = w^\dagger M w + w^\dagger M' w \ge 0$. Also, $w^\dagger M^* w = ((w^*)^\dagger M w^*)^*
\ge 0$. Finally, $Re(M) = (M+M^*)/2$ which is positive semidefinite according to the previous cases.
\end{proof}

\begin{claim}\label{claim:psd_det}
The determinant of a positive semidefinite matrix $M$ is non-negative.
\end{claim}
\begin{proof}
Since $M$ is Hermitian, it can be diagonalized with orthogonal eigenvectors
 and real eigenvalues. Moreover, since it is
positive semidefinite, its eigenvalues are non-negative. Hence, the
 determinant of $M$, which is the product of its
eigenvalues, is non-negative.
\end{proof}

Next, we define a positive definite function over an arbitrary group $E$.
In this paper, $E$ will always be a grid in
$\R^n$, i.e., a discrete additive subgroup of $\R^n$.

\begin{definition}\label{def:positive_definite}
Let $E$ be a group. A function $g:E \rightarrow \C$ is positive
definite (PD) if for any integer $k \ge 1$ and any set
of group elements $x_1,\ldots,x_k \in E$, the $k$ by $k$ matrix
 $M$ defined by $M_{i,j}=g(x_i-x_j)$ is positive semidefinite.
\end{definition}

The following two corollaries follow directly from
 Definition \ref{def:positive_definite} and Claims
\ref{claim:psd_simple}, \ref{claim:psd_det}:

\begin{corollary}\label{cor:pd_simple}
Let $g,g'$ be two positive definite functions. Then the following functions are also positive definite: $c\cdot g$,
$g+g'$, $Re(g)$ where $c>0$ is real.
\end{corollary}

\begin{corollary}\label{cor:pd_determinant}
Let $g:E \rightarrow \C$ be a positive definite function for some group $E$. Then, for any integer $k \ge 1$ and any
set of group elements $x_1,\ldots,x_k \in E$, the $k$ by $k$ matrix $M$ defined by $M_{i,j}=g(x_i-x_j)$ has a non-negative
determinant.
\end{corollary}

Using Corollary~\ref{cor:pd_determinant} we derive the following two useful lemmas. These lemmas describe known
properties of positive definite functions (see, e.g., \cite{Sasvari,Schlather}).

\begin{lemma}\label{pdf2x2}
Let $g:E \rightarrow \R$ be a real positive definite function such that $g(0)=1$. Then for any
$x\in E$, $g(x)=g(-x)$ and $|g(x)| \le 1$.
\end{lemma}
\begin{proof}
Choose $k=2$ in Definition \ref{def:positive_definite} and choose $0$ and $x$ as the two group elements. Then,
$$ M = \left(%
 \begin{array}{cc}
  1 & g(x) \\
  g(-x) & 1 \\
 \end{array}%
 \right)
$$
is positive semidefinite. Hence, $M$ is
 Hermitian and $g(x)=(g(-x))^*=g(-x)$. Moreover,
$$ 0\le \left|%
M \right|=
 \left|%
 \begin{array}{cc}
  1 & g(x) \\
  g(x) & 1 \\
 \end{array}%
 \right|= 1-(g(x))^2
$$
Therefore,
$$ |g(x)| \le 1. $$
\end{proof}

\begin{lemma}\label{pdf3x3}
Let $g:E \rightarrow \R$ be a real positive definite function such that $g(0)=1$. Then, for any $x\in E$ such that $x/2
\in E$ exists, $g(x/2) \le \sqrt{(1+g(x))/2} \le (g(x)+3)/4$.
\end{lemma}
\begin{proof}
Choose $k=3$ in Definition \ref{def:positive_definite} and choose
 $0$, $x$ and $x/2$ as
the three group elements.
 Let $b$ denote $g(x)$ and $a$ denote $g(x/2)=g(-x/2)$. Then,
\begin{eqnarray*}
0 \le \left|%
\begin{array}{ccc}
  1 & b & a \\
  b & 1 & a \\
  a & a & 1 \\
\end{array}%
 \right|=  1-a^2 - b(b-a^2) + a(ba-a)=(1-b)(1+b-2a^2).
 \end{eqnarray*}
According to Lemma~\ref{pdf2x2}, $b\le 1$. Hence we have $1+b-2a^2 \ge 0 $
which implies
\begin{eqnarray*}
a \le \sqrt{(1+b)/2} \le \frac{3+b}{4}.
\end{eqnarray*}
\end{proof}

\subsection{Autocorrelation and Positive Definite Functions}

The following claim shows the important fact that autocorrelation
functions are always positive definite.

\begin{claim}\label{cl:pd}
Let $f$ be a function from a group $E$ to the complex numbers, and let $h$ be its autocorrelation function defined by
$h(x) := \sum_{y\in E} f^*(y) f(y+x).$ Then $h$ is a positive definite function.
\end{claim}

\begin{proof}
Let $k\ge 1$ and $x_1,\ldots,x_k \in E$ be arbitrary and consider the $k\times k$ matrix $M$ defined by
$M_{i,j}=h(x_i-x_j)$. According to Definition~\ref{def:positive_definite}, it is enough to
 show that $M$ is PSD. For
any vector $w\in \C^k$,
\begin{eqnarray*}
 w^\dag M w &=& \sum_{i,j=1}^k h(x_i-x_j) w^*_i w_j
             =  \sum_{i,j=1}^k \sum_{y\in E} f^*(y) f(y+x_i-x_j) w^*_i w_j \\
            &=& \sum_{i,j=1}^k  \sum_{y\in E} f^*(y-x_i) f(y-x_j) w^*_i w_j
             =  \sum_{y\in E}( \sum_{i=1}^k f^*(y-x_i)w^*_i)(\sum_{j=1}^k f(y-x_j)w_j)  \\
            &=& \sum_{y\in E} \left| \sum_{i=1}^k f(y-x_i)w_i \right|^2\ge 0
\end{eqnarray*}
\end{proof}

\section{QMA+}\label{sec:qma+}

A ``super-verifier'' is given by a classical polynomial-time
randomized algorithm that given an input $x$ outputs a
description of a quantum circuit $V$ and two numbers $r,s \in [0,1]$.
This can be thought of as follows. Assume that we
are given a witness described by a density matrix $\rho$. Then, consider $\tr(\Pi^{\ket{1}}V \rho V^\dag)$ where
$\Pi^{\ket{1}}$ is the projection on the space where the output qubit of $V$ is $\ket{1}$ (this is equal to the
probability of measuring an output qubit of $\ket{1}$). Then, $r$ represents an estimate of this value and $s$ is the
accuracy of the estimate.

\begin{definition}[QMA+] \label{Def:QMA+}
A language $L \in QMA+$ if there exists a super-verifier and polynomials $p_1,p_2,p_3$ such that:
\begin{itemize}
 \item $\forall x \in L \quad \exists \rho \quad
           \Pr_{V,r,s} \left(  | \tr(\Pi^{\ket{1}} V \rho V^\dag) - r | \le s \right)=1$ \\
         (i.e., there exists a witness such that with probability $1$ the super-verifier outputs $V$
         which accepts the witness with probability which is close to $r$)
 \item $\forall x \notin L \quad \forall \rho \quad
           \Pr_{V,r,s} \left( | \tr(\Pi^{\ket{1}} V \rho V^\dag) - r | \le s+p_3(1/|x|) \right)\leq 1-p_2(1/|x|)$ \\
         (i.e., for any witness, with some non-negligible probability, the super-verifier outputs a circuit $V$
         that accepts the witness with probability which is not close to $r$)
\end{itemize}
where probabilities are taken over the outputs $V,r,s$ of the super-verifier and $\rho$ is a density matrix over
$p_1(|x|)$ qubits.
\end{definition}

In the rest of this section we prove Theorem \ref{thm:qma+}. We note that
for simplicity we defined $QMA+$ with
perfect completeness in the YES case; the same
 theorem holds also with non-perfect completeness.

The following lemma proves the easy direction of the theorem. It will not be used in this paper and is presented here
mainly for the sake of completeness.
\begin{lemma}
$QMA \subseteq QMA+$
\end{lemma}
\begin{proof}
Note that using amplification \cite{KitaevBook}, any language in $QMA$ has a verifier with completeness $c \ge 7/8$ and
soundness $s \le 1/8$. Given such a verifier $V$, construct a super-verifier that simply outputs $(V,r=1,s=1/2)$. This
satisfies the definition of $QMA+$, using $p_3(|x|)=p_2(|x|)=1/4$, for example.
\end{proof}

We now prove the more interesting direction:

\begin{theorem}\label{thm:inqma}
$QMA+ \subseteq QMA$
\end{theorem}
\begin{proof}
Given a super-verifier for a language $L\in QMA+$ with
polynomials $p_1,p_2,p_3$, we construct a QMA verifier $V'$ for
$L$. Let $k=poly(|x|)$ be a large enough
parameter to be determined later. The witness given to $V'$ consists of $k
\cdot p_1(|x|)$ qubits which can be thought of as
$k$ registers of $p_1(|x|)$ qubits each. Given an input $x$, the
verifier $V'$ starts by calling the super-verifier with the input $x$.
 The result is a description of a circuit $V$ and
numbers $r,s\in [0,1]$. Next, $V'$ applies
$V$ to each of the $k$ registers and measures the results. Let $r'$ denote
the number of 1s measured divided by $k$.
 $V'$ accepts if $|r'-r| \le s + \frac{1}{2}p_3(1/|x|)$ and rejects otherwise.

{~}

\noindent\underline{{\bf Completeness: }} Let $x\in L$ and let $\rho$ be as
in Definition \ref{Def:QMA+}. The witness for $V'$ will be
$\rho^{\otimes k}$. Note that the probability to measure $1$ in each register
is  $\tr(\Pi^{\ket{1}} V
\rho V^\dag)$. Let us denote this probability by $p_V$ and
let us choose $k = n / (p_3(1/|x|))^2$. Then, according to
the Chernoff bound, the probability that $|r'-p_V| >
\frac{1}{2}p_3(1/|x|)$ is at most $2 e^{-2k (p_3(1/|x|)/2)^2} = 2^{-\Omega(n)}$.
By Definition \ref{Def:QMA+}, the triples $(V,r,s)$ given by the
 super-verifier are such that  $|p_V - r| \le s$ and
$$|r'-r| \le |r' - p_V| + |p_V - r| \le\frac{1}{2}p_3(1/|x|)+s$$
which implies that $V'$ accepts with probability exponentially close
to $1$.

{~}

\noindent\underline{{\bf  Soundness:}}
It suffices to show that if $x\notin L$ then $V'$ rejects
with probability
at least $\frac{1}{2}p_2(1/|x|) p_3(1/|x|)$ (which is polynomially
bounded from $0$). Essentially, the reasoning is based on a Markov argument, as we will see
shortly.

Let $|\eta\ra$ be any witness for $V'$.  We
 first define a witness $\rho$ for the circuits $V$ that
the super-verifier outputs.
 Let $\eta_i$ be the reduced density matrix
of $\eta$ to the $i$'th register, and let $\rho$ to be the average of the
reduced density matrices: $\rho=\frac{1}{k} \sum_{i=1}^k \eta_i$.
For an output of the super-verifier $(V,r,s)$ we again
let $p_V$ denote the probability to measure $1$ given $\rho$,
namely  $p_V=\tr(\Pi^{\ket{1}} V \rho V^\dag)$. We observe that
\begin{claim}\label{claim:exp}
For a fixed witness $|\eta\ra$ and a fixed circuit $V$, the expectation of the
random variable $r'$ is $p_V$.
\end{claim}
\begin{proof}
The random variable $r'$ is the average of $k$ indicator variables. The expected value
of the $i$'th indicator variable is $\tr(\Pi^{\ket{1}} V \eta_i V^\dag)$. Therefore,
using linearity of expectation, the expected value of $r'$ is
$ \frac{1}{k} \sum \tr(\Pi^{\ket{1}} V \eta_i V^\dag) = p_V$.
\end{proof}

According to Definition \ref{Def:QMA+}, with probability at least $p_2(1/|x|)$,
$(V,r,s)$ is such that
$| p_V - r | > s+ p_3(1/|x|)$.
Then, it is enough to show that for such triples $(V,r,s)$, $V'$
rejects with probability at least $\frac{1}{2} p_3(1/|x|)$.
So, in the following fix one such triple $(V,r,s)$.
Using Claim \ref{claim:exp}, we obtain that the expected value of $r'$ is
either less than $r-s-p_3(1/|x|)$ or more than $r+s+p_3(1/|x|)$.
We now use a Markov argument;
In the first case, since $r'$ is a non-negative random variable, the
probability that it is more than $r-s-\frac{1}{2}p_3(1/|x|)$ (so that $V'$ may accept) is at most
$$ \frac{r-s - p_3(1/|x|) } {r-s- \frac{1}{2} p_3(1/|x|)}
   \le 1 - \frac{1}{2} p_3(1/|x|).$$
Similarly, for the second case, consider the non-negative random variable
$1 - r'$. The probability that it is greater than $1 - (r + s + \frac{1}{2}p_3(1/|x|))$ is at most
$$ \frac{1 - (r+s+p_3(1/|x|))}{1 - (r + s + \frac{1}{2}p_3(1/|x|))} \le 1 - \frac{1}{2}p_3(1/|x|).$$
\end{proof}

\section{coGapCVP' is in QMA+}\label{sec:cvp}

In this section we prove Theorem \ref{thm:gapcvp'}. Recall that an input to $coGapCVP'_{c\sqrt{n}}$ is a pair $(L,v)$.
By choosing a large enough constant $c$ and scaling we can assume that in $YES$ instances, $d(v,L)> 10\sqrt{n}$ and the
shortest vector in $L$ is of length at least $10\sqrt{n}$ and that in $NO$ instances $d(v,L)\le 1/3$.

\subsection{The Quantum Witness}\label{sec:quantum_witness}

In the case of a $YES$ instance, the prover provides a quantum state that represents a Gaussian distribution around the
lattice points. We will use the periodicity of the lattice and present our state as a superposition over points inside
the parallelepiped $\calP(L)$.

We would have liked to consider the superposition over all points in the parallelepiped $\calP(L)$ with weights that
depend on the distance to the lattice:
\begin{eqnarray*}
\ket{\xi} \approx \sum_{x\in \calP(L) ~|~ d(x,L) \le 2\sqrt{n} } \sqrt{\mu(\tau_L(x))} \ket{x}.
\end{eqnarray*}
However, this state is ill defined since the register
contains points in $\R^n$, which we need infinite precision in
order to represent. We will therefore discretize space, and consider points on a
very fine lattice $G$. In order to prevent confusion, we will refer to $G$ as a `grid' and not a lattice.
We discuss this in the following.

\begin{description}\label{sec:disc}
\item {\bf Discretization Issues:}
The grid $G$ is obtained by scaling down the lattice $L=(v_1,\ldots,v_n)$ by a factor of
$2^m$ for some $m>0$. Formally, $G$ is the set
of all integer combinations of the vectors $v_i/2^m$ where $m \le poly(n)$
 is chosen such that the following
requirements are satisfied:

~~~~~~$\bullet$ The diameter of one parallelepiped of $G$, $\diam(\calP(G))$, is at most $2^{-n^2}$, and

~~~~~~$\bullet$ $m\ge \ell+n$ where $\ell$ was defined as the precision in which $v$ is given.

Note that we can choose $m$ to be polynomial in $n$ because $\diam(\calP(G))=\diam(\calP(L))/2^m \le \sum_i |v_i|/2^m$.

To store a vector in $\calP(L)\cap G$ in the quantum register, we store its coefficients in terms of the basis vectors
$v_i$. Each coefficient is a number of the form $j/2^m$ for $0\le j < 2^m$
and so we need $m$ bits to store $j$.
 Since we need $n$ coefficients,
 the register consists of $nm = poly(n)$ qubits.
\end{description}

\noindent The formal definition of the witness is:

\begin{eqnarray*}
\ket{\xi}= \sum_{x\in \calP(L)\cap G} f(x) \ket{x}
\end{eqnarray*}
where
\begin{eqnarray*}
f(x)=\left\{%
\begin{array}{ll}
    \sqrt{\mu(\tau_L(x))}/D  \quad & d(x,L) \le 2  \sqrt{n}, \\
     0 & \hbox{otherwise.} \\
\end{array}%
\right.
\end{eqnarray*}
and $D$ is a normalization factor chosen so that
$$ \sum_{x\in \calP(L)\cap G} (f(x))^2 = 1.$$

\subsection{Autocorrelation tests}

Our verification process is based on autocorrelation tests which we define in the following.

\begin{deff} For $x\in G$,
$T_x$ is defined to be the bijection $a \mapsto a-x ~ mod~ \calP(L)$ from $\calP(L)\cap G$ into itself.
\end{deff}
\begin{deff}
The function $g: G \rightarrow \R$ is defined as $ g(x) = Re(\la \xi| T_x |\xi\ra) $.
\end{deff}
Note that $g(x)$ is equal to
\begin{eqnarray*}
g(x)= \sum_{y\in \calP(L)\cap G} f(y) f(x+y)
\end{eqnarray*}

\begin{definition}[Autocorrelation circuit with respect to $x$]\label{def:1bit}
For any $x\in G$ define the circuit $C_x$ as follows. Given an input register, add one qubit (called the control qubit)
in the state $\frac{1}{\sqrt{2}}(\ket{0}+\ket{1})$. Then apply $T_x$ to the register conditioned that the control qubit
is $1$, and otherwise do nothing. Finally, apply the Hadamard matrix $H$ on the control qubit. The control qubit is the
output qubit.
\end{definition}

\begin{claim}\label{claim:purestate_correlation}
Given a pure state $\ket{\eta}$, the probability of measuring 1 after applying $C_x$ is $(1-Re(\bra{\eta}T_x
\ket{\eta}))/2$.
\end{claim}
\begin{proof}
After adding the control qubit to $\ket{\eta}$, the state is $\frac{1}{\sqrt{2}}(\ket{0}\ket{\eta}+\ket{1}\ket{\eta})$.
After performing a conditioned $T_x$, the state is $\frac{1}{\sqrt{2}}(\ket{0}\ket{\eta}+\ket{1}T_x\ket{\eta})$.
Finally, after the Hadamard transform, the state is
 \begin{eqnarray*}
 \frac{1}{2} \left(  \ket{0}(\ket{\eta}+T_x\ket{\eta})+ \ket{1}(\ket{\eta}-T_x\ket{\eta})  \right) .
 \end{eqnarray*}
The probability of measuring $1$ is therefore
 \begin{eqnarray*}
 \frac{1}{4}(\la \eta | - \la \eta |T_x^\dag )(\ket{\eta} - T_x\ket{\eta}) =
  \frac{1}{2} (1-Re(\la\eta|T_x|\eta\ra)).
 \end{eqnarray*}
\end{proof}

The next lemma provides a good approximation to $g(x)$:
\begin{lemma}\label{le:integrals}
Let $L$ be a lattice whose shortest vector is
of length at least $10\sqrt{n}$. Then, for any $x\in G$,
$$ \left| g(x) - \mu(\tau_L(x)/2) \right| \le 2^{-\Omega(n)}.$$
\end{lemma}

\begin{proof}
The proof is fairly complicated technically,
and we delay it to the appendix.
\end{proof}

\subsection{The super-verifier}
The super-verifier randomly chooses one of the following two cases:
\begin{itemize}
\item {\bf Autocorrelation with respect to $v$}

Output the circuit $C_v$, together with $r=1/2$ and $s=n^{-100}$.

\item {\bf Autocorrelation with respect to short vectors}

Let $B'$ denote the ball of radius $n^{-10}+n^{-11}$ around the origin.
 Choose a vector $x \in B' \cap G$ from the uniform distribution over $B' \cap G$. Let $x'$ be either $x$ or $2x$ with equal probability. Output the circuit
$C_{x'}$, together with $r=(1-\mu(x'/2))/2$ and $s=n^{-100}$.

\end{itemize}

\subsection{Efficiency of the verifier}
The verifier works on points in $\calP(L)\cap G$. Note that the map $a \mapsto a - x ~ mod ~\calP(L)$ for $x\in \calP(L)
\cap G$ is well defined and is a bijection on $\calP(L) \cap G$, and so is its inverse. This means that these maps can
be applied efficiently by a quantum computer. This follows from a basic result in quantum computation, which states
that if $U$ and its inverse can be applied efficiently classically, then they can be applied efficiently and without
garbage bits by a quantum computer \cite{KitaevBook}.

 Next, we describe a procedure that picks a point
 uniformly at random from $B' \cap G$. First, pick a point $z\in
\R^n$ uniformly from the ball $(n^{-10}+n^{-11}+n^{-20})B_n$. Represent
 it as a combination of the basis vectors
$v_1,...,v_n$. Then, let $x\in G$ be the point obtained by rounding the coefficients of $z$ down to multiples of
$2^{-m}$. If $x\in B'$ then output $x$. Otherwise, repeat the procedure again.

The probability of outputting each $x\in B' \cap G$ is proportional to the probability that $z$ is in $x+\calP(G)$.
Since $\diam(\calP(G)) < n^{-20}$, $x+\calP(G) \subseteq (n^{-10}+n^{-11}+n^{-20})B_n$ and therefore the above
probability is proportional to the volume of $\calP(G)$. This volume is the same for all $x$ and hence the output is
indeed uniform over $B' \cap G$. The procedure has to be repeated when $x\notin B'$. This can only happen if $\|z\| \ge
n^{-10}+n^{-11}-\diam(\calP(G))>n^{-10}+n^{-11}-n^{-20}$. But the probability of this is at most
$$ 1 - \left( \frac{n^{-10}+n^{-11}-n^{-20}}{n^{-10}+n^{-11}+n^{-20}} \right) ^n =
   1 - \left( 1 - 2 \frac{n^{-20}}{n^{-10}+n^{-11}+n^{-20}} \right) ^n \le 1-(1-2n^{-10})^n \le 2n^{-9}$$
and therefore the procedure stops after a polynomial number of steps with probability exponentially close to 1.
Finally, we note that we cannot really choose a uniform point $z$ in the ball since its representation is not finite;
this can be easily fixed by choosing an approximation to $z$ and then arguing that the distance of the
output distribution from the uniform distribution on $B' \cap G$ is exponentially small.

\subsection{Completeness}

\begin{claim}
Let $L$ be a lattice whose shortest vector is of length at least $10\sqrt{n}$ and $v$ a vector such that $d(v,L)\ge
10 \sqrt{n}$. Then, given the witness $\ket{\xi}$ described in Section \ref{sec:quantum_witness}, the super-verifier
outputs triples $(V,r,s)$ such that $| \tr(\Pi^{\ket{1}} V \ket{\xi}\bra{\xi} V^\dag) - r | \le s$.
\end{claim}
\begin{proof}
First assume that the super-verifier outputs $C_v$. By Lemma \ref{le:integrals}, $g(v)$ is exponentially small and
therefore, using Claim \ref{claim:purestate_correlation}, $\tr(\Pi^{\ket{1}} V \ket{\xi}\bra{\xi} V^\dag )=(1-g(v))/2$
is in the range $[\frac{1}{2}-n^{-100}, \frac{1}{2}+n^{-100}]$. Otherwise, the super-verifier outputs a circuit
$C_{x'}$ for some short vector $x'$. Notice that $d(x',L)=\|x'\|$, since the lattice has no short vectors. By Lemma
\ref{le:integrals}, $g(x')$ is exponentially close to $\mu(x'/2)$ and hence $\tr(\Pi^{\ket{1}} V \ket{\xi}\bra{\xi}
V^\dag)$ is exponentially close to $(1-\mu(x'/2))/2$.
\end{proof}

\subsection{Soundness}

\begin{theorem}
Let $L$ be a lattice and $v$ be a vector such that $d(v,L)\le 1/3$. Then, given any witness $\rho$, with probability
at least $n^{-1000}$, the super-verifier outputs triples $(V,r,s)$ such that $| \tr(\Pi^{\ket{1}} V \rho V^\dag) - r | > s$.
\end{theorem}

\begin{proof}
We will need the following definitions:

\begin{definition}
We say $x$ is ``good'' for a real function $h$ if $|h(x)- \mu(x/2) | \le 2n^{-100}$ and $|h(2x)- \mu(x) | \le
2n^{-100}$. Otherwise, we say $x$ is ``bad'' for $h$.
\end{definition}

\begin{definition}
We say that $h$ is $\epsilon$-Gaussian approximating on
the set $A$ if all except at most $\epsilon$ fraction of the vectors in
$A$ are good for $h$.
\end{definition}

The idea of the proof is as follows. Let $\rho$ be any witness and assume by contradiction that $d(v,L)\le 1/3$ and
that with probability at least $1-n^{-1000}$ the super-verifier outputs $(V,r,s)$ such that $| \tr(\Pi^{\ket{1}} V \rho
V^\dag) - r | \le s$. We use $\rho$ to define a PD function $h$, and show that by the conditions of the theorem
$|h(v)|\le 2n^{-100}$ and that $h$ is $n^{-200}$-Gaussian approximating on $B'\cap G$. We then show that such a PD
function doesn't exist, if $d(v,L)\le 1/3$, which derives a contradiction.

\paragraph{Definition of $h$:}
We can write $\rho$ as $\rho = \sum_i w_i \ket{\alpha_i} \bra{\alpha_i}$ for some weights $w_i$ and pure states
$\ket{\alpha_i}$. Also, write $\ket{\alpha_i} = \sum_{y\in \calP(L) \cap G} \beta_i(y) \ket{y}$ for some
$\beta_i:\calP(L) \cap G \rightarrow \C$. This form of $\ket{\alpha_i}$ is without loss of generality, because by our
choice of the number of qubits in the register, and by the definition of $G$, each possible basis state represents a
point in $\calP(L) \cap G$.
 Define the functions $h_i:G \rightarrow \C$,
$$ h_i(x) = \bra{\alpha_i} T_x \ket{\alpha_i} = \sum_{y\in \calP(L) \cap G} \beta_i^*(y) \beta_i (y+x ~\mod~ \calP(L) ).$$
We let $h: G \rightarrow \R$ be the function
$$ h(x) = \sum_i w_i Re(h_i(x)).$$

\begin{claim}
$h$ is PD.
\end{claim}

 \begin{proof}
According to Corollary \ref{cor:pd_simple}, it is enough to show that the $h_i$'s are positive definite. This follows
from Claim \ref{cl:pd}, using the group of points in $\calP(L) \cap G$ with addition modulo $\calP(L)$.\end{proof}

\begin{claim}
$|h(v)|\le 2n^{-100}$.
\end{claim}

\begin{proof}
The super-verifier outputs the triple $(C_v, \frac{1}{2}, n^{-100})$ with probability half. By the assumption of the
theorem we thus know that
\begin{equation}\label{eq:1}
 | \tr(\Pi^{\ket{1}} C_v \rho C_v^\dag) - \frac{1}{2} | \le n^{-100}.
\end{equation}
Note that by Claim \ref{claim:purestate_correlation},
for any $x \in G$,
\begin{equation}\label{eq:tr}
\tr(\Pi^{\ket{1}} C_x \rho C_x^\dag )=\sum_i
w_i(1-Re(h_i(x)))/2 = (1-h(x))/2.
\end{equation}
Substituting equation (\ref{eq:tr}) in (\ref{eq:1}) and multiplying by 2 we get,
$$ | h(v) | \le 2n^{-100}.$$
\end{proof}

\begin{claim}
$h$ is $n^{-200}$-Gaussian approximating on the set $B'\cap G$.
\end{claim}

\begin{proof}
The super-verifier outputs a triple of the form $(C_{x'}, (1-\mu(x'/2))/2, n^{-100})$ with probability half. Hence,
with probability at least $1-2n^{-1000}$
\begin{equation}\label{eq:2}
 | \tr(\Pi^{\ket{1}} C_{x'} \rho C_{x'}^\dag ) - (1-\mu(x'/2))/2 | \le n^{-100}
\end{equation}
where the probability is taken on the choice of $x'$ by the super-verifier. Substituting equation (\ref{eq:tr}) in
equation (\ref{eq:2}) and multiplying by 2, we get:
$$ | h(x') - \mu(x'/2) | \le 2n^{-100}.$$
Recall that $x'$ is chosen in two steps: we first choose $x \in B' \cap G$ and then choose $x'$ to be either $x$ or
$2x$. Hence, with probability at least $1-4n^{-1000}$ over the choice of $x$, both
$$ | h(x) - \mu(x/2) | \le 2n^{-100}$$
and
$$ | h(2x) - \mu(x) | \le 2n^{-100}$$
hold. Hence, $h$ is $n^{-200}$-Gaussian approximating on the set $B' \cap G$.
\end{proof}

We obtain a contradiction by using the following lemma with $w=\tau_L(v)$. Recall that the coefficients of $v$ in the
lattice basis are multiples of $2^{-\ell}$. This implies that $\tau_L(v)$ can be represented as an integer combination
of the vectors $v_i/2^\ell$. Since $m$ was chosen to be at least $\ell+n$, $w/2^n \in G$.
 The proof of the lemma appears
in the next section.

\begin{lemma}\label{le:noconditions}
Let $w \in G$ such that $w/2^n$ is also in $G$ and $\|w\| \le 1/3$. Then, there is no positive definite function $h$,
$h(0)=1$, which is $n^{-200}$-Gaussian approximating on $B'\cap G$ and $|h(w)|\le 2n^{-100}$.
\end{lemma}
\end{proof}

\subsection{Proof of Lemma \ref{le:noconditions}: No such PD function}
\begin{proof} (Of Lemma \ref{le:noconditions})
Assume by contradiction that $h$ is a positive definite function, that $|h(w)| \le 2n^{-100}$ and that $h$ is
$n^{-200}$-Gaussian approximating on $B' \cap G$. We will derive a contradiction in two steps. First, we will find a
short vector $y$ in $w$'s direction such that $h(y)$ is much lower than the Gaussian value of $\mu(y/2)$. This is done
using the upper bound on $|h(w)|$ and ``pulling'' it towards the origin using the PD conditions. We will then apply a
lemma that shows that the same deviation from the Gaussian occurs everywhere and not only in $w$'s direction.

\begin{deff}
Define $y=w/2^k$, where $k \ge 0$ is the minimal integer such that $\|y\| \le 2n^{-12}$.
\end{deff}

Notice that if $k\neq 0$ then $\|y\| > n^{-12}$. Hence, using $\|w\| \le 1/3$, we get that $k \le \log (n^{12}/3)$.

\begin{claim}
 $y \in G$.
\end{claim}
\begin{proof}
Since $k < n$, $y$ is an integer multiple of $w/2^n$ and is therefore in $G$.
\end{proof}

The Gaussian at $y$, $\mu(y/2)$, can be approximated by $1-\frac{\pi}{4}\|y\|^2$ which is at least $1-\frac{\pi}{4}
4n^{-24}= 1-\pi n^{-24}$. The following claim shows that $h(y)$ is strictly less than the Gaussian at $y$:

\begin{claim} Let $h$ be PD such that $h(0)=1$, $h(w)\le n^{-100}$ and $\|w\| \le 1/3$.
Then $h(y)\le 1-5 n^{-24}$.
\end{claim}

\begin{proof}
Lemma~\ref{pdf3x3}, using $w$, $w/2$, shows that $h(w/2) \le \frac{3+2n^{-100}}{4}$. Applying Lemma~\ref{pdf3x3} again
gives $h(w/4) \le \frac{15+2n^{-100}}{16}$, and applying it $k$ times gives $h(y)=h(w/2^k)\le
1-\frac{1-2n^{-100}}{2^{2k}}$. Since $k \le \log (n^{12}/3)$, we have $h(y)\le 1-\frac{1-2n^{-100}}{2^{2k}} \le
1-\frac{1-2n^{-100}}{\frac{1}{9} n^{24}} \le  1-5n^{-24}$.
\end{proof}

To derive a contradiction, we will use the following claim:

\begin{claim}\label{cl:psdapprox}
Let $h$ be PD such that $h(0)=1$ and $h(y)\le 1-5 n^{-24}$ for some $\|y\|\le 2n^{-12}$. Let $z\in G$ be such that $\|
P_y(z)\| \le 1/n^{100}$ and $|\|P_{y^\bot}(z)\|-n^{-10}|\le n^{-100}$. Then at least one of the vectors $z,z-y,z+y$ is
bad for $h$.
\end{claim}

\begin{proof}
The proof uses the PD condition with $4\times 4$ matrices. It is quite technical, and is delayed to the appendix.
\end{proof}

We want to show that in the verifier's second test, it has a non negligible chance of picking $x$ which is equal to one
of the vectors of the form $z,z+y,z-y$ satisfying the requirements in Claim \ref{cl:psdapprox}.  This would mean it has
a good chance of catching a ``bad'' vector, as we will see later. For this we define:
\begin{eqnarray*}
 A_1 &=& \left\{ z\in \R^n ~|~ |\|P_{y^\bot}(z)\|-n^{-10}|\le n^{-100} ~and~ \|P_{y}(z)\| \le n^{-100} \right\} \\
 A_2 &=& A_1 + y\\
 A_3 &=& A_1 - y\\
\end{eqnarray*}

\begin{claim}
$A_1,A_2, A_3\subseteq B'$
\end{claim}
\begin{proof}
By the triangle inequality, the norm of a vector in $A_1,A_2$ or $A_3$ is at most $n^{-10}+2n^{-100}+2n^{-12}$, because
$\|y\|\le 2n^{-12}$. Hence, its norm is less than $n^{-10}+n^{-11}$, the radius of $B'$.
\end{proof}

\begin{claim}
Let $x$ be chosen uniformly at random from $B'\cap G$. The probability for $x$ to be in $A_i\cap G$ is at least
$n^{-180}/10$, for all $i=1,2,3$.
\end{claim}
\begin{proof}
First notice that $|A_1\cap G|=|A_2\cap G|=|A_3\cap G|$ and they are all subsets of $B'$. Hence, the probability that
$x$ is $A_i\cap G$ is the same for $i=1,2,3$. Therefore, in the following it will be enough to consider the set $A_1$.
Let
$$ \tilde{A} = \left\{ z\in \R^n ~|~ |\|P_{y^\bot}(z)\|-n^{-10}|\le n^{-100}/2 ~and~ \|P_{y}(z)\| \le n^{-100}/2 \right\}$$
be a subset of $A_1$. Notice that since $\diam(\calP(G))$ was chosen to be very small, any point $z \in G$ such that
$(z+\calP(G)) \cap \tilde{A} \neq \phi$ must satisfy $z \in A_1$. Similarly, if we define $\tilde{B}$ as the ball of
radius $n^{-10}+2n^{-11}$ then any point $z\in G$ such that $(z+\calP(G)) \cap B' \neq \phi$ must satisfy $z \in
\tilde{B}$. Hence we obtain,
$$ \frac{|A_1 \cap G|}{|B' \cap G|} \ge \frac{ \vol(\tilde{A})/\vol(\calP(G))}{ \vol(\tilde{B})/\vol(\calP(G))} =
   \frac{ \vol(\tilde{A})}{ \vol(\tilde{B})}. $$

We now lower bound this ratio of volumes. Recall that the volume of an $n$ dimensional ball around the origin of radius
$R$ is $\omega_nR^n$ where $\omega_n$ is the volume of the unit $n$-ball.
\begin{eqnarray*}
\vol(\tilde{A}) &=& n^{-100} \cdot \omega_{n-1}\cdot ( (n^{-10}+n^{-100}/2)^{n-1} - (n^{-10}-n^{-100}/2)^{n-1}) \\
         &\ge& n^{-100} \cdot \omega_{n-1}\cdot (n^{-10}-n^{-100}/2)^{n-1} \cdot ( (1+n^{-90})^{n-1} - 1) \\
         &\ge& n^{-100} \cdot \omega_{n-1}\cdot (n^{-10}-n^{-100}/2)^{n-1} \cdot n^{-90}.
\end{eqnarray*}
Using $\vol(\tilde{B})=\omega_n \cdot (n^{-10}+2n^{-11})^n$,
\begin{eqnarray*}
 \frac{\vol(\tilde{A})}{\vol(\tilde{B})} &\ge& n^{-190} \cdot  \frac{\omega_{n-1}}{\omega_n}
      \cdot \frac{(n^{-10}-n^{-100}/2)^{n-1}}{(n^{-10}+2n^{-11})^n} \\
      &\ge& n^{-180} \cdot  \frac{\omega_{n-1}}{\omega_n}
      \cdot \frac{(1-n^{-90}/2)^{n-1}}{(1+2n^{-1})^n} \ge n^{-180}/10
\end{eqnarray*}
where in the last inequality we used $\omega_{n-1}/\omega_n = \Omega(\sqrt{n}) > 1$.
\end{proof}

\begin{claim}
$h$ is not $n^{-200}$-Gaussian approximating on $B'\cap G$.
\end{claim}
\begin{proof}
For any $x\in A_1 \cap G$, consider the triple $x,x+y,x-y$ and notice that $x+y \in A_2 \cap G$, $x-y \in A_3 \cap G$.
By Claim \ref{cl:psdapprox}, at least one point in each triple is bad for $h$. Hence, at least a third of the points in
one of the sets $A_1\cap G,A_2 \cap G,A_3 \cap G$ are bad for $h$. Since each of these sets contains $n^{-180}/10$ of
the points in $B' \cap G$, the fraction of bad points for $h$ in $B'\cap G$ is at least $n^{-180}/30$.
\end{proof}
This is a contradiction and thus
 completes the proof of Lemma \ref{le:noconditions}.
\end{proof}

\section{Reducing coGapSVP to coGapCVP'}\label{sec:svptocvp}

In this section we prove Theorem \ref{thm:cvptosvp}. We show how to construct a verifier $V'$ for $coGapSVP_\beta$
given a verifier $V$ for $coGapCVP'_{\beta}$. By using amplification \cite{KitaevBook},
 we can assume without loss of generality that for $YES$ instances
 there exists a witness such
that $V$ accepts with probability at least $1-2^{-n}$ and that for $NO$ instances $V$ accepts with probability less
than $2^{-n}$ for any witness. Let $L$ be the input lattice given by $(v_1,\ldots,v_n)$. The witness supplied to $V'$
is supposed to be of the following form:
$$ \ket{\alpha_1} \ket{\alpha_2} \ldots \ket{\alpha_n}.$$
Each $\ket{\alpha_i}$ is supposed to be a witness for the $coGapCVP'_{\beta}$ instance given by the lattice $L_i$
spanned by $(v_1,\ldots,v_{i-1},2v_i,v_{i+1},\ldots,v_n)$ and the target vector $v_i$. The verifier $V'$ applies $V$ to
each $\ket{\alpha_i}$ with the instance $(L_i,v_i)$. It accepts if and only if $V$ accepted in all the calls.

First assume that $L$ is a $YES$ instance to $coGapSVP_{\beta}$. In other words, the length of the shortest vector is
at least $\beta$.
Since $L_i$ is a sublattice of $L$,
its shortest vector is at least $\beta$.
In addition, since for any $i\in [n]$, $v_i\notin L_i$
this implies that $d(v_i,L_i) \ge \beta$. Hence, $(L_i,v_i)$
 is a $YES$ instance of
$coGapCVP'_{\beta}$ and there exists a witness $\ket{\alpha_i}$ such that $V$ accepts it with probability at least
$1-2^{-n}$. Therefore, the combined witness $\ket{\alpha_1} \ldots \ket{\alpha_n}$ is accepted by $V'$ with probability
at least $1-n2^{-n}$.

It is left to consider the case where $L$ is a $NO$ instance. In other words, if
$$u=a_1 v_1+a_2 v_2 + \ldots + a_n v_n$$
denotes the shortest vector, then its length is at most $1$. Notice that not all the $a_i$'s are even for otherwise the
vector $u/2$ is a shorter lattice vector. Let $j$ be such that $a_j$ is odd. Then the distance of $v_j$ from the
lattice $L_j$ is at most $\|u\| \le 1$ since $v_j + u \in L_j$. Hence, the $j$'th instance of $coGapCVP'_{\beta}$ is a
$NO$ instance and for any witness $\ket{\alpha_j}$, $V$ accepts with probability at most $2^{-n}$ and so does $V'$.

\section{Acknowledgments}
We would like to thank Hartmut Klauck, Alex Samordnitsky, Benny Sudakov, Umesh Vazirani and John Watrous for helpful
discussions. OR thanks Martin Schlather for sending a copy of his technical report.

\bibliographystyle{plain}

\begin{thebibliography}{10}

\bibitem{DoritQMASurvey}
D.~Aharonov and T.~Naveh.
\newblock Quantum {NP} - a survey.
\newblock In {\em quant-ph/0210077, http://xxx.lanl.gov}, 2002.

\bibitem{AharonovTaShma}
D.~Aharonov and A.~Ta-Shma.
\newblock Quantum adiabatic state generation and stataistical zero knowledge.
\newblock In {\em Proc. 35th ACM Symp. on Theory of Computing}, San Diego, CA,
  June 2003.

\bibitem{adiabatic}
D.~Aharonov, W.~van Dam, J.~Kempe, Z.~Landau, S.~Lloyd, and O.~Regev.
\newblock On the universality of quantum adiabatic computation on a {2D}
  lattice (temporary name).
\newblock 2003.
\newblock Manuscript.

\bibitem{ChildsSpeedupWalk}
A.~M. Childs, R.~Cleve, E.~Deotto, E.~Farhi, S.~Gutmann, and D.~A. Spielman.
\newblock Exponential algorithmic speedup by quantum walk.
\newblock In {\em Proc. 35th ACM Symp. on Theory of Computing}, San Diego, CA,
  June 2003.

\bibitem{EttingerHoyer}
M.~Ettinger and P.~H{\o}yer.
\newblock On quantum algorithms for noncommutative hidden subgroups.
\newblock {\em Advances in Applied Mathematics}, 25(3):239--251, 2000.

\bibitem{santha}
K.~Friedl, G.~Ivanyos, F.~Magniez, M.~Santha, and P.~Sen.
\newblock Hidden translation and orbit coset in quantum computing.
\newblock In {\em Proc. 35th ACM Symp. on Theory of Computing}, 2003.

\bibitem{GG}
O.~Goldreich and S.~Goldwasser.
\newblock On the limits of nonapproximability of lattice problems.
\newblock {\em J. Comput. System Sci.}, 60(3):540--563, 2000.

\bibitem{GMWGraphIsomorphism}
O.~Goldreich, S.~Micali, and A.~Wigderson.
\newblock Proofs that yield nothing but their validity, or {A}ll languages in
  {NP} have zero-knowledge proof systems.
\newblock {\em J. Assoc. Comput. Mach.}, 38(3):691--729, 1991.

\bibitem{GMSS}
O.~Goldreich, D.~Micciancio, S.~Safra, and J.-P. Seifert.
\newblock Approximating shortest lattice vectors is not harder than
  approximating closest lattice vectors.
\newblock {\em Inform. Process. Lett.}, 71(2):55--61, 1999.

\bibitem{vazirani}
M.~Grigni, L.~Schulman, M.~Vazirani, and U.~Vazirani.
\newblock Quantum mechanical algorithms for the non-abelean hidden subgroup
  problem.
\newblock In {\em Proc. 33th ACM Symp. on Theory of Computing}, pages 68--74,
  2001.

\bibitem{hallgren}
S.~Hallgren.
\newblock Polynomial-time quantum algorithms for {P}ell's equation and the
  principal ideal problem.
\newblock In {\em Proc. 34th ACM Symp. on Theory of Computing}, pages 653--658,
  2002.

\bibitem{KempeRegev}
J.~Kempe and O.~Regev.
\newblock 3-local hamiltonian is {QMA}-complete.
\newblock In {\em quant-ph/0302079, http://xxx.lanl.gov}, 2003.

\bibitem{KitaevBook}
A.~Yu. Kitaev, A.~H. Shen, and M.~N. Vyalyi.
\newblock {\em Classical and quantum computation}, volume~47 of {\em Graduate
  Studies in Mathematics}.
\newblock AMS, 2002.

\bibitem{LagariasKZ}
J.~C. Lagarias, H.~W. Lenstra, Jr., and C.-P. Schnorr.
\newblock Korkin-{Z}olotarev bases and successive minima of a lattice and its
  reciprocal lattice.
\newblock {\em Combinatorica}, 10(4):333--348, 1990.

\bibitem{Micciancio01hnf}
D.~Micciancio.
\newblock Improving lattice based cryptosystems using the hermite normal form.
\newblock In {\em Cryptography and Lattices Conference (CaLC)}, volume 2146 of
  {\em Lecture Notes in Computer Science}, pages 126--145, Providence, Rhode
  Island, March 2001. Springer-Verlag.

\bibitem{MicciancioBook}
D.~Micciancio and S.~Goldwasser.
\newblock {\em Complexity of Lattice Problems: a cryptographic perspective},
  volume 671 of {\em The Kluwer International Series in Engineering and
  Computer Science}.
\newblock Kluwer Academic Publishers, Boston, Massachusetts, March 2002.

\bibitem{RegevQuantumLattices}
O.~Regev.
\newblock Quantum computation and lattice problems.
\newblock In {\em Proceedings of the 43rd Annual Symposium on Foundations of
  Computer Science (FOCS) 2002}, Vancouver, Canada, November 2002.

\bibitem{RegevPKE}
O.~Regev.
\newblock New lattice based cryptographic constructions.
\newblock In {\em Proc. 35th ACM Symp. on Theory of Computing}, San Diego, CA,
  June 2003.

\bibitem{Sasvari}
Z.~Sasv{\'a}ri.
\newblock {\em Positive definite and definitizable functions}, volume~2 of {\em
  Mathematical Topics}.
\newblock Akademie Verlag, Berlin, 1994.

\bibitem{Schlather}
M.~Schlather.
\newblock Introduction to positive definite functions and to unconditional
  simulation of random fields.
\newblock Technical report ST 99-10, Lancaster University, 1999.

\bibitem{ShorFactor}
P.~W. Shor.
\newblock Polynomial-time algorithms for prime factorization and discrete
  logarithms on a quantum computer.
\newblock {\em {SIAM Journal on Computing}}, 26(5):1484--1509, 1997.

\bibitem{ShpilkaRaz}
A.~Shpilka and R.~Raz.
\newblock On the power of quantum proofs.
\newblock 2002.
\newblock Unpublished.

\bibitem{legendre}
W.~van Dam, S.~Hallgren, and L.~Ip.
\newblock Quantum algorithms for some hidden shift problems.
\newblock In {\em Proceedings of the ACM-SIAM Symposium on Discrete
  Algorithms}, pages 489--498, 2003.

\bibitem{watrous}
J.~Watrous.
\newblock Succinct quantum proofs for properties of finite groups.
\newblock In {\em Proceedings of the 41st Annual Symposium on Foundations of
  Computer Science}, pages 537--546, 2000.

\end{thebibliography}

\begin{appendix}

\section{Some Technical Claims}

\begin{claim}\label{cl:mu_error}
For any two vectors $z,z' \in \R^n$,
$$ |\mu(z)-\mu(z')| \le O(\|z-z'\|).$$
\end{claim}
\begin{proof}
The derivative of $\mu(\alpha)$ is $ -2\pi\alpha e^{-\pi \alpha^2} $ which is at most $\sqrt{2 \pi / e}$ in absolute
value. Hence, for any $\alpha,\beta\in \R$,
$$ |\mu(\alpha)-\mu(\beta)| \le \sqrt{2 \pi / e} \cdot |\alpha-\beta| = O(|\alpha-\beta|).$$
The claim follows since for any $w\in \R^n$, $\mu(w)= \mu(\|w\|)$ and $|\|z\|-\|z'\|| \le \|z-z'\|$.
\end{proof}

\begin{claim}\label{cl:int_mu}
$$\int_{\R^n} \mu(z) dz = 1$$
\end{claim}
\begin{proof}
$$\int_{\R^n} \mu(z) dz = \int_{\R^n} e^{-\pi \|z\|^2} dz =
  \int_{\R^n} e^{-\pi z_1^2} \cdot \ldots \cdot e^{-\pi z_n^2} dz = (\int_\R e^{-\pi x^2} dx)^n = 1^n = 1.$$
\end{proof}

\begin{claim}\label{cl:tail_mu}
$$\int_{\sqrt{n} B_n} \mu(z) dz \ge 1 - 2^{-\Omega(n)}$$
\end{claim}
\begin{proof}
According to Claim \ref{cl:int_mu}, it is enough to show that
$$\int_{\R^n \setminus \sqrt{n} B_n} \mu(z) dz \le 2^{-\Omega(n)}.$$
Since $\mu$ depends only on the norm of $z$ we can switch to polar coordinates and get
\begin{eqnarray}
 && n \cdot \omega_n \cdot \int_{\sqrt{n}}^\infty e^{-\pi r^2} r^{n-1} dr \le  \nonumber \\
 && 2n \cdot \omega_n \cdot \int_{\sqrt{n}}^\infty e^{-\pi r^2} r^{n-1} (1 - \frac{n-2}{2\pi r^2}) dr = \nonumber \\
 && 2n \cdot \omega_n \cdot \left. \left( -\frac{1}{2\pi} e^{-\pi r^2} r^{n-2} \right) \right|^{\infty}_{\sqrt{n}} =
      \nonumber\\
 && \frac{n}{\pi} \cdot \omega_n \cdot e^{-\pi n} n^{\frac{n}{2}-1}. \label{eq:tail_mu_1}
\end{eqnarray}
Using Stirling's formula,
$$ \omega_n = \frac{\pi^{n/2}}{\Gamma(\frac{n}{2}+1)} \approx \frac{1}{\sqrt{\pi n}} (\frac{2\pi e}{n})^{n/2}.$$
Hence, (\ref{eq:tail_mu_1}) is
$$ \frac{n}{\pi} \cdot \frac{1}{\sqrt{\pi n}} \cdot (\frac{2\pi e}{n})^{n/2} e^{-\pi n} n^{\frac{n}{2}-1}
   = \frac{1}{\pi} \cdot \frac{1}{\sqrt{\pi n}} \cdot (2\pi e)^{n/2} e^{-\pi n} = 2^{-\Omega(n)}.$$
\end{proof}

\section{Proof of correct autocorrelation}

In this section we prove Lemma \ref{le:integrals}. Recall that $g$ is defined as
$$ g(x) = \sum_{y\in \calP(L)\cap G} f(y) f(x+y).$$

The function $f$ is periodic on the lattice $L$. Hence,
$$ g(x) = \sum_{y\in \calP(L)\cap G} f(y) f(x+y) =
   \sum_{y\in \calP(L)\cap G} f(\tau_L(y)) f(\tau_L(y)+x).$$
Furthermore, $\tau_L$ can be seen as a bijection between $\calP(L)\cap G$ and $Vor(L) \cap G$. Hence, the above is
equal to,
$$\sum_{y\in Vor(L) \cap G} f(y) f(y+x).$$
When $\|y\| > 2\sqrt{n}$, $f(y)=0$. Also, if $\|y\| \le 2\sqrt{n}$ then $y\in Vor(L)$ because the shortest vector in
the lattice is at least $10\sqrt{n}$. Therefore, the above sum is,
$$\sum_{y\in G  ~|~ \|y\| \le 2\sqrt{n}} f(y) f(y+x).$$
Notice that for $\|y\| \le 2\sqrt{n}$, $f(y) = \sqrt{\mu(y)}/D$. Also, if $f(y+x)\neq 0$ then $d(y+x,L) \le 2\sqrt{n}$
and therefore $d(x,L) \le 4\sqrt{n}$. Using the assumption that the shortest vector in the lattice is $10\sqrt{n}$,
this implies that the closest lattice point to $y+x$ is the same as the closest lattice point to $x$. In other words,
$\tau_L(y+x)=y+\tau_L(x)$. Let $S(x)$ denote the set of all $y\in G$ such that both $\|y\|$ and $\|y+\tau_L(x)\|$ are
at most $2\sqrt{n}$. Then the above sum is,
$$\frac{1}{D^2} \sum_{y\in S(x)} \sqrt{\mu(y) \mu(y+\tau_L(x))}.$$

For any $y\in S(x)$, $\mu(y) \ge 2^{-O(n)}$. Using Claim \ref{cl:mu_error}, we see that for any $z\in y+\calP(G)$,
$|\mu(y)-\mu(z)|\le O(\diam(\calP(G))) = 2^{-\Omega(n^2)}$. Hence, this translates to a multiplicative error of $\mu(z)
= (1 \pm 2^{-\Omega(n)})\mu(y)$. A similar argument shows that $\mu(z+\tau_L(x)) = (1 \pm
2^{-\Omega(n)})\mu(y+\tau_L(x))$. By combining the two equalities and taking the square root, we get that for any $y\in
S(x)$ and for any $z \in y+\calP(G)$,
$$ \sqrt{\mu(y)\mu(y+\tau_L(x))} = (1 \pm 2^{-\Omega(n)}) \sqrt{\mu(z)\mu(z+\tau_L(x))}.$$
Averaging the right hand side over all $z \in y+\calP(G)$,
$$ \sqrt{\mu(y)\mu(y+\tau_L(x))} = (1 \pm 2^{-\Omega(n)}) \frac{1}{\vol(\calP(G))} \int_{y+\calP(G)} \sqrt{\mu(z)\mu(z+\tau_L(x))} dz.$$
We therefore obtain the following estimation of $g(x)$:
\begin{eqnarray*}
 &&(1\pm 2^{-\Omega(n)}) \frac{1}{\vol(\calP(G)) \cdot D^2} \sum_{y\in S(x)} \int_{y+\calP(G)}
    \sqrt{\mu(z)\mu(z+\tau_L(x))} dz \\
 && \quad = (1\pm 2^{-\Omega(n)}) \frac{1}{\vol(\calP(G)) \cdot D^2} \int_{S(x)+\calP(G)} \sqrt{\mu(z)\mu(z+\tau_L(x))} dz.
\end{eqnarray*}

Recall that $D$ was chosen so that $g(0)=1$. Hence, we get that
$$ (1\pm 2^{-\Omega(n)}) \frac{1}{\vol(\calP(G)) \cdot D^2} \int_{S(0)+\calP(G)} \mu(z) dz = 1.$$
Since $S(0)+\calP(G)$ contains the ball of radius $\sqrt{n}$ around the origin,
$$ 1 - 2^{-\Omega(n)} \le \int_{\sqrt{n} B_n} \mu(z) dz \le \int_{S(0)+\calP(G)} \mu(z) dz \le \int_{\R^n} \mu(z) dz = 1$$
where we used Claim \ref{cl:tail_mu} and Claim \ref{cl:int_mu}. Hence,
$$ \frac{1}{\vol(\calP(G))\cdot D^2} = 1 \pm 2^{-\Omega(n)}.$$

Thus, the estimation of $g(x)$ becomes
$$ (1 \pm 2^{-\Omega(n)}) \int_{S(x)+\calP(G)} \sqrt{\mu(z)\mu(z+\tau_L(x))} dz.$$
This can be further approximated by
\begin{eqnarray*}
 && (1 \pm 2^{-\Omega(n)}) \int_{S(x)+\calP(G)} \sqrt{\mu(z)\mu(z+\tau_L(x))} dz = \\
 && (1 \pm 2^{-\Omega(n)}) \int_{S(x)+\calP(G)} \mu(z+\tau_L(x)/2) \mu(\tau_L(x)/2) dz = \\
 && (1 \pm 2^{-\Omega(n)}) \mu(\tau_L(x)/2) \int_{S(x)+\calP(G)} \mu(z+\tau_L(x)/2) dz.
\end{eqnarray*}
where in the first equality we used $\|z\|^2+\|z+\tau_L(x)\|^2=2(\|z+\tau_L(x)/2\|^2 + \|\tau_L(x)/2\|^2)$.

We can now upper bound $g(x)$ by
$$ (1 \pm 2^{-\Omega(n)}) \mu(\tau_L(x)/2) \int_{\R^n} \mu(z+\tau_L(x)/2) dz =
   (1 \pm 2^{-\Omega(n)}) \mu(\tau_L(x)/2) \int_{\R^n} \mu(z) dz = (1 \pm 2^{-\Omega(n)}) \mu(\tau_L(x)/2).$$
In particular, this means that for $x$ such that $d(x,L)$ is greater than, say, $\sqrt{n}/2$, $g(x)$ is indeed
exponentially close to $\mu(\tau_L(x)/2) = 2^{-\Omega(n)}$. Therefore, it remains to consider the case $d(x,L) \le
\sqrt{n}/2$. Here, $\sqrt{n} B_n \subseteq S(x)+\calP(G)+\tau_L(x)/2$ and therefore $g(x)$ can be lower bounded by
 $$(1 \pm 2^{-\Omega(n)}) \mu(\tau_L(x)/2) \int_{S(x)+\calP(G)+\tau_L(x)/2} \mu(z) dz \ge
   (1 \pm 2^{-\Omega(n)}) \mu(\tau_L(x)/2) \int_{\sqrt{n} B_n} \mu(z) dz \ge (1 \pm 2^{-\Omega(n)}) \mu(\tau_L(x)/2)
 $$
where we used Claim \ref{cl:tail_mu}.

\section{Proof of Claim \ref{cl:psdapprox}}

We assume by contradiction that the vectors $z,z-y,z+y$ are good for $h$, that $h(y)\le 1-5 n^{-24}$ and that $h$ is a
positive definite function. We will derive a contradiction by using the PD condition with a $4\times 4$ matrix.

Choose $k=4$ in Definition \ref{def:positive_definite} and choose the origin, the vector $-z$, the vector $z$ and the
vector $y$ as the four vectors. By the assumption that $h$ is positive definite, and by Corollary
\ref{cor:pd_determinant}, the following holds:

\begin{eqnarray*}
\left|
\begin{array}{cccc}
  1    & h(z)   & h(z)   & h(y) \\
  h(z) & 1      & h(2z)  & h(z+y) \\
  h(z) & h(2z)  & 1      & h(z-y) \\
  h(y) & h(z+y) & h(z-y) & 1 \\
\end{array}
 \right| \ge 0.
\end{eqnarray*}

By the assumption that $z,z-y,z+y$ are good, it follows that
\begin{eqnarray*}
 h(z) &=& \mu(z/2) + O(n^{-100}) \\
 h(2z) &=& \mu(z) + O(n^{-100}) \\
 h(z+y) &=& \mu((z+y)/2) + O(n^{-100}) \\
 h(z-y) &=& \mu((z-y)/2) + O(n^{-100})
\end{eqnarray*}
where the $O(n^{-100})$ denotes an additive error whose absolute value is at most in the order of $n^{-100}$.

Let $z'=P_{y^\bot}(z)$ be the projection of $z$ on the subspace orthogonal to $y$. According to Claim
\ref{cl:mu_error}, by replacing $z$ with $z'$ in the above estimations we introduce an error of at most $O(\|z-z'\|)
\le O(n^{-100})$:
\begin{eqnarray*}
 h(z) &=& \mu(z'/2) + O(n^{-100}) \\
 h(2z) &=& \mu(z') + O(n^{-100}) \\
 h(z+y) &=& \mu((z'+y)/2) + O(n^{-100}) \\
 h(z-y) &=& \mu((z'-y)/2) + O(n^{-100})
\end{eqnarray*}

Let $\alpha = \mu(z'/2)$ and $\beta = \mu(z'/2)\mu(y/2)$. Then, notice that $\mu(z')=\alpha^4$ and that
$\mu((z'+y)/2)=\mu((z'-y)/2)=\beta$ since $z'$ and $y$ are orthogonal. Hence,
\begin{eqnarray*}
 h(z) &=& \alpha + O(n^{-100}) \\
 h(2z) &=& \alpha^4 + O(n^{-100}) \\
 h(z-y) &=& \beta + O(n^{-100}) \\
 h(z+y) &=& \beta + O(n^{-100})
\end{eqnarray*}

We can replace each entry of the above determinant by
its estimation. By Lemma \ref{pdf2x2}, all the entries of the
determinant have an absolute value of at most one and therefore
the error introduced is at most $O(n^{-100})$:

\begin{eqnarray*}
\left|
\begin{array}{cccc}
  1 & \alpha & \alpha & h(y) \\
  \alpha & 1 & \alpha^4 & \beta \\
  \alpha & \alpha^4 & 1 & \beta \\
  h(y) & \beta & \beta & 1 \\
\end{array}
 \right| + O(n^{-100}) \ge 0.
\end{eqnarray*}

Let us now expand the determinant:
\begin{eqnarray*}
\left|
\begin{array}{cccc}
  1 & \alpha & \alpha & h(y) \\
  \alpha & 1 & \alpha^4 & \beta \\
  \alpha & \alpha^4 & 1 & \beta \\
  h(y) & \beta & \beta & 1 \\
\end{array}
 \right| &=&
 \left|
\begin{array}{cccc}
  1 & \alpha               & \alpha               & h(y) \\
  0 & 1- \alpha^2          & \alpha^4 - \alpha^2 & \beta - \alpha h(y) \\
  0 & \alpha^4 - \alpha^2 & 1 - \alpha^2         & \beta - \alpha h(y) \\
  0 & \beta - \alpha h(y)  & \beta - \alpha h(y)  & 1 -  (h(y))^2 \\
\end{array}
 \right| =
  \\
 \left|
\begin{array}{ccc}
  1- \alpha^2          & \alpha^4 - \alpha^2 & \beta - \alpha h(y) \\
  \alpha^4 - \alpha^2 & 1 - \alpha^2         & \beta - \alpha h(y) \\
  \beta - \alpha h(y)  & \beta - \alpha h(y)  & 1 -  (h(y))^2 \\
\end{array}
 \right| &=&
 \left|
\begin{array}{ccc}
  1- \alpha^2          & \alpha^4 - \alpha^2 & \beta - \alpha h(y) \\
  \alpha^4 - 1 & 1 - \alpha^4         & 0 \\
  \beta - \alpha h(y)  & \beta - \alpha h(y)  & 1 -  (h(y))^2 \\
\end{array}
 \right| =
 \\
 \left|
\begin{array}{ccc}
  1- \alpha^2          & (\alpha^2 - 1)^2 & \beta - \alpha h(y) \\
  \alpha^4 - 1 & 0         & 0 \\
  \beta - \alpha h(y)  & 2(\beta - \alpha h(y))  & 1 -  (h(y))^2 \\
\end{array}
 \right| &=&
(1-\alpha^4) \left( (\alpha^2-1)^2 (1 - (h(y))^2) - 2 (\beta - \alpha h(y))^2 \right) .
\end{eqnarray*}
Hence,
$$ (1-\alpha^4) \left( (\alpha^2-1)^2 (1 - (h(y))^2) - 2 (\beta - \alpha h(y))^2 \right) + O(n^{-100}) \ge 0.$$
From the assumption that $|\|z'\|-n^{-10}| \le n^{-100}$ it follows that $1-\alpha^4$ is in the order of $O(n^{-20})$.
Hence, dividing by $1-\alpha^4$ which is a positive number, we get
$$ (\alpha^2-1)^2 (1 - (h(y))^2) - 2 (\beta - \alpha h(y))^2 + O(n^{-80}) \ge 0.$$
Rearranging terms,
$$ \left( (1-\alpha^2)^2 - 2 \beta^2 \right) + 4\alpha \beta \cdot h(y) - (1+\alpha^4) \cdot (h(y))^2 + O(n^{-80}) \ge 0.$$
From this we can obtain the following lower bound on $h(y)$,
$$ h(y) \ge \frac{2\alpha\beta - \sqrt{4\alpha^2\beta^2 + (1+\alpha^4)((1-\alpha^2)^2-2\beta^2)+O(n^{-80})}}{1+\alpha^4}.$$
We show that the term under the square root is negligible, because it is $O(n^{-64})$:
\begin{eqnarray*}
 4\alpha^2\beta^2 + (1+\alpha^4)((1-\alpha^2)^2 - 2\beta^2) + O(n^{-80}) &=&
   (1-\alpha^2)^2 (1+\alpha^4-2\beta^2) + O(n^{-80}) \\
   &=&(1-\alpha^2)^2 \left( ( 1 - \alpha^2)^2 + 2 (\alpha - \beta)(\alpha+\beta) \right) + O(n^{-80}).
\end{eqnarray*}
The term $1-\alpha^2$ is of the order $O(n^{-20})$, the term $\alpha+\beta$ is at most $2$ and the term $\alpha-\beta$
equals $\mu(z'/2)(1-\mu(y/2)) \le 1-\mu(y/2)$ which is of the order $O(n^{-24})$. Hence, the above expression is of the
order $O(n^{-64})$. After taking the square root it is of the order $O(n^{-32})$. We therefore get:
$$ h(y) \ge \frac{2\alpha\beta}{1+\alpha^4}+O(n^{-32}) = \frac{2\alpha^2}{1+\alpha^4}\cdot \frac{\beta}{\alpha}+O(n^{-32}).$$
We have
$$ \frac{2\alpha^2}{1+\alpha^4} = 1 - O((\alpha-1)^2) = 1 + O(n^{-40}),$$
where we used the Taylor series expansion of the left hand side around $1$.

Also, $\beta/\alpha = \mu(y/2) \ge 1 - \pi n^{-24} + O(n^{-48})$. Hence,
$$ h(y) \ge 1 - \pi n^{-24} + O(n^{-32})$$
which contradicts the assumption that $h(y) \le 1-5 n^{-24}$.

\end{appendix}
\end{document}